\documentclass
[showkeys,superscriptaddress,showpacs,preprintnumbers,byrevtex]{revtex4} 
\usepackage{amsmath,amsfonts,amssymb,amscd,amsxtra,amsthm}
\usepackage{graphicx}
\usepackage{epstopdf}
\usepackage{bm}
\begin{document}   
\preprint{CYCU-HEP-10-18}
\title{QCD under magnetic field: chiral magnetic effect at heavy ion collisions}      
\author{Seung-il Nam}
\email[E-mail: ]{sinam@kau.ac.kr}
\affiliation{Research Institute of Basic Sciences, Korea Aerospace University, Goyang, 412-791, Korea} 
\author{Chung-Wen Kao}
\email[E-mail: ]{cwkao@cycu.edu.tw}
\affiliation{Department of Physics, Chung-Yuan Christian University, Chung-Li 32023, Taiwan}
\author{Byung-Geel Yu}
\email[E-mail: ]{bgyu@kau.ac.kr}
\affiliation{Research Institute of Basic Sciences, Korea Aerospace University, Goyang, 412-791, Korea} 
\date{\today}
\begin{abstract}
In this talk, we report our recent studies for the chiral magnetic effect, which signals the $P$- and $CP$-violations at heavy ion collisions. We compute the electric current and its correlations, induced by the external magnetic field, inside the hot and dense QCD matter created in the HIC. For this purpose, we employ the instanton-liquid model, modified by the Harrington-Shepard caloron at finite $T$. We observe that the chiral magnetic effect current and its correlations increase with respect to the magnetic field, whereas decease as  functions of $T$, due to the diluting instanton ensemble. It turns out that the numerical results are in good agreement with those from the model-independent analyses and lattice QCD simulations. We also reproduce the charge separations, observed in the STAR experiment, qualitatively well, considering the simplified Li\'enard-Wiechert potential, screening and size effects.  
\end{abstract}  
\maketitle
\section{Introduction}
It has been well known that the {\it quantum chromodynamics} (QCD) is the first principle for the strongly interacting systems, consisting of quarks and gluons. Recent energetic developments in heavy ion collisions (HIC), such as the relativistic heavy ion collider (RHIC) and large hadron collider (LHC), starts to shed the light on the new path for studying QCD. Due to the nontrivial structure of the QCD vacuum, this fundamental theory enjoys the richness of the various  phases, characterized by the relevant symmetries and invariances, and their breakdowns at finite temperature ($T$) and/or density ($\mu$). Thus, studies on this QCD phase structure will reveal the profound understandings for the symmetries as well as QCD vacuum itself. In this sense, the hot and dense QCD matter, i.e. {\it quark gluon plasma} (QGP), believed to be created through the HIC experiments, is the right place to investigate these interesting subjects in QCD.

Recently, among the various experimental and theoretical subjects in the HIC experiments, the $P$ violation, indicated by the charge separation, has been attracting interest greatly~\cite{Kharzeev:2004ey,Voloshin:2004vk,Kharzeev:2007jp,Voloshin:2008jx,:2009txa,:2009uh,Fukushima:2008xe,Fukushima:2009ft,Warringa:2009rw,Asakawa:2010bu,Buividovich:2009wi,Buividovich:2010tn,Nam:2009jb,Nam:2009hq,Nam:2010nk}. This interesting phenomena is called the chiral magnetic effect (CME). Microscopically, under the strong external magnetic field $\bm{B}$, generated through the peripheral HIC, the CME can be interpreted as the induced electric current, due to the $CP$ violation from the nontrivial QCD vacuum structure, indicted by the nonzero topological charge $Q_\mathrm{t}\propto N_L-N_R$, according to the axial U(1) anomaly. Hence, nonzero $Q_\mathrm{t}$ generates chirality inside the QCD matter. In this sense, observing the $P$ violation experimentally, one can find the signals for the $P$ as well as $CP$ violations in the HIC experiments. 

Taking into account that the $Q_\mathrm{t}$ is also proportional to the number difference of the instantons and anti-instantons, $\Delta=N_I-N_{\bar{I}}$, we are motivated to study the CME, employing the liquid-instanton model with nonzero $\Delta$. The magnetic field is included using the linear Schwinger method~\cite{Schwinger:1951nm}. To extend the model to a finite-$T$ system, we modify the model with the Harrington-Shepard caloron~\cite{Harrington:1976dj} and fermionic Matsubara formula. We compute the CME current, its correlations, and charge separations as numerical results, and compare them with theory and experiment. This talk is based on our recent works~\cite{Nam:2009nn,Nam:2009jb,Nam:2009hq,Nam:2010nk}, and readers may find more details on this talk in them and references therein.

The present report is organized as follows: In Section II, we introduce the theoretical formalism briefly. The analytical and numerical results are given in Section III with discussions. Final Section is devoted to summary and outlook.     
\section{Formalism}
\begin{figure}[t]
\begin{tabular}{cc}
\includegraphics[width=8.5cm]{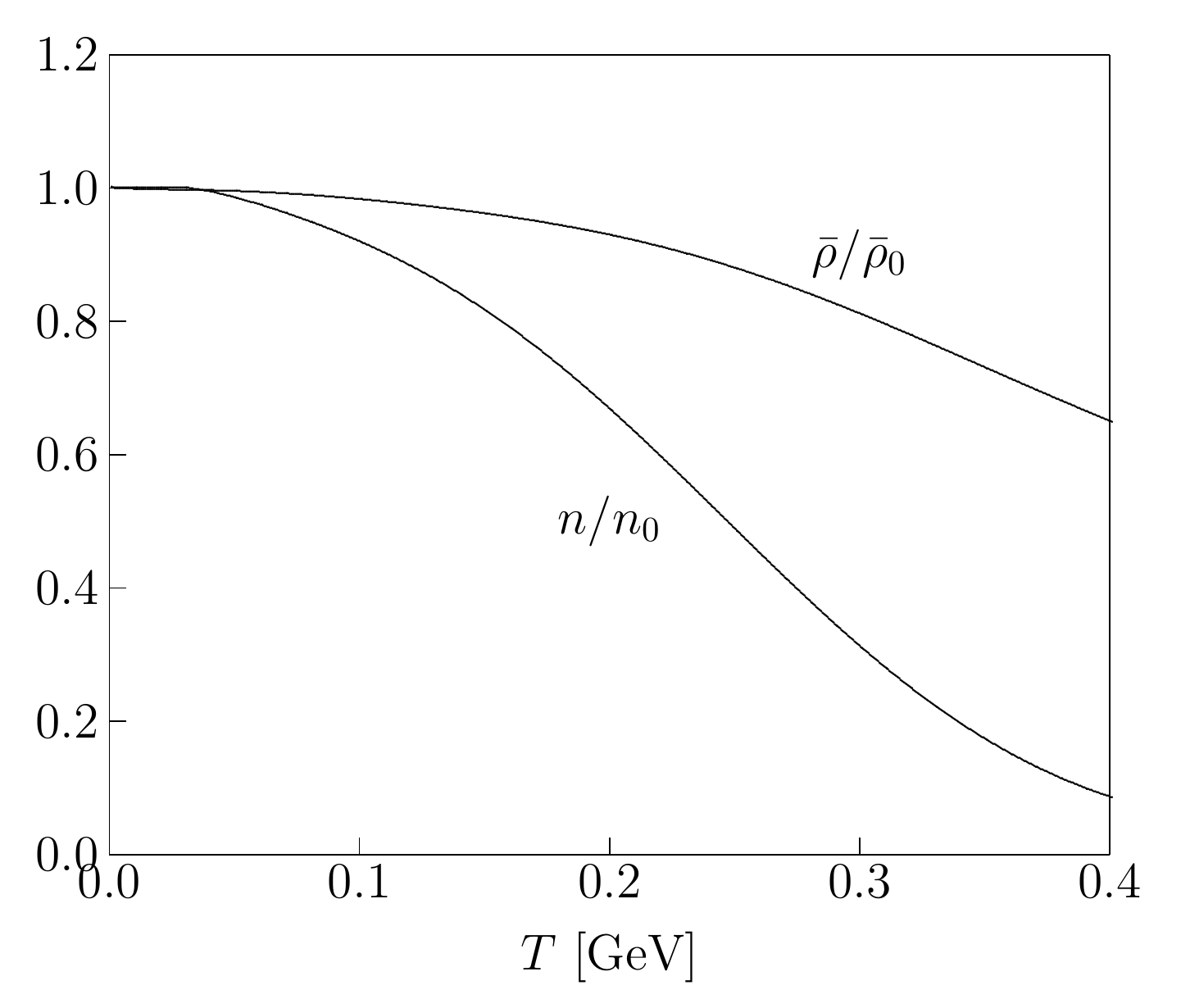}
\includegraphics[width=8.5cm]{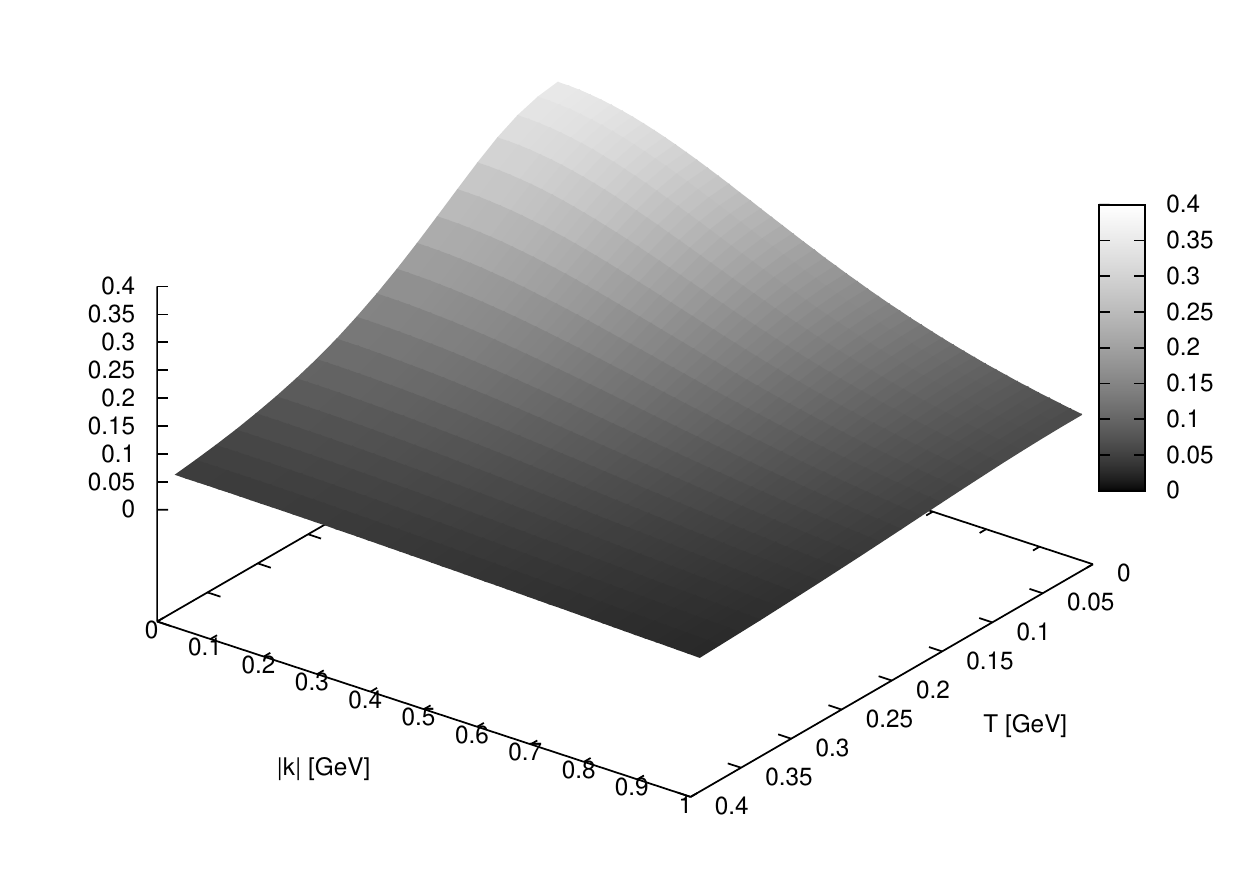}
\end{tabular}
\caption{Instanton parameters, $\bar{R}$ and $\bar{\rho}$  as functions of $T$ (left) and the three-momentum and $T$ dependent constituent-quark mass $M$ (right).}       
\label{FIG1n2}
\end{figure}

In order to study the CME and related quantities, we write the effective action, derived from the liquid-instanton vacuum with $P$ and $CP$ violations~\cite{Diakonov:1995qy,Nam:2010nk} as follows:
\begin{eqnarray}
\label{eq:EA}
\mathcal{S}_{\mathrm{eff}}
&=&-\int\frac{d^4k}{(2\pi)^4}\mathrm{Tr}_{c,f,\gamma}\ln
\left[\rlap{\,/}{K}-i(1+\delta\gamma_{5})M+\rlap{/}{V}\right],
\end{eqnarray}
where $K$ and $M$ denote the quark four momentum $K=k+A+V$ and momentum-dependent constituent-quark mass, respectively. $\delta$ and $V$ stand for the strength of the $CP$ violation, being proportional to the nontrivial topological charge $Q_\mathrm{t}$, and external vector source field. Detailed interpretations for deriving Eq.~(\ref{eq:EA}) are given in Refs.~\cite{Nam:2009jb,Nam:2010nk}. Using the effective action, the vacuum expectation value (VEV) of the electric current can be obtained by the functional differentiation of the $\mathcal{S}_{\mathrm{eff}}$ with respect to the external vector source field $V_\mu$ in a standard way. In order to include the magnetic field, which corresponds to that generated through the HIC, we use the linear Schwinger method~\cite{Schwinger:1951nm}. To this end, we expand the quark propagator in the presence of the external electromagnetic (EM) field.  Plugging this EM field modified quark propagator into the CME currents, we then expand it with respect to the field strength tensor and collect the terms proportional to the magnetic field.  Being similarly to the CME current, their correlations can be easily derived using generic functional methods. See Refs.~\cite{Nam:2009jb,Nam:2009hq,Nam:2010nk} for more details. All the physical quantities, CME current and correlations, are extended to those at finite $T$. For this purpose,  we employ the Harrington-Shepard caloron~\cite{Harrington:1976dj}, which is a classical solution of the Yang-Mill's equation being periodic in the temporal direction. Using this nontrivial solution, one can modify the instanton parameters, such as the average inter-(anti)instanton distances $\bar{R}$ and their sizes $\bar{\rho}$. In Fig.~\ref{FIG1n2}, we depict the instanton parameters, $\bar{R}$ and $\bar{\rho}$  as functions of $T$ (left) and the three-momentum and $T$ dependent constituent-quark mass $M$ (right). As shown in the left panel, the instanton ensemble becomes diluting as $T$ increases. As a consequence, the qaurk-instanton interaction is diminished as $T$ grows, resulting in decreasing constituent-quark mass as shown in the right panel of Fig.~\ref{FIG1n2}.  At the same time, it plays the role of the natural regulator, which tames the UV divergence appearing in the loop integrals. 
In addition to this modification, the integrals appearing in the relevant physical quantities are replaced by the fermionic Matsubara formula~\cite{Nam:2009nn}.  For reproducing the charge separation, which is suggested to be the signal of the $P$ and $CP$ violations and reported to be measured by STAR collaboration at RHIC, one has to take into account three ingredients: 1) spatial configuration of the magnetic field~\cite{Kharzeev:2007jp}, 2) screening effect caused by the nontrivial interactions inside the QGP, resulting in that the CME current is mainly produced on the surface of the QGP~\cite{Kharzeev:2007jp}, and 3) size effect gives a factor $1/N^2_{\mathrm{nucl}}$ where $N_{\mathrm{nucl}}$ stands for the atomic mass of the projectile nucleus~\cite{Kharzeev:2007jp,Nam:2010nk}. As for the magnetic field, we consider the simplified Li\'enard-Wiechert potential~\cite{Nam:2010nk}. Again, more details for computing the charge separations can be found in Ref.~\cite{Nam:2010nk}. 
\section{analytical and numerical results}
\begin{figure}[t]
\includegraphics[width=8.5cm]{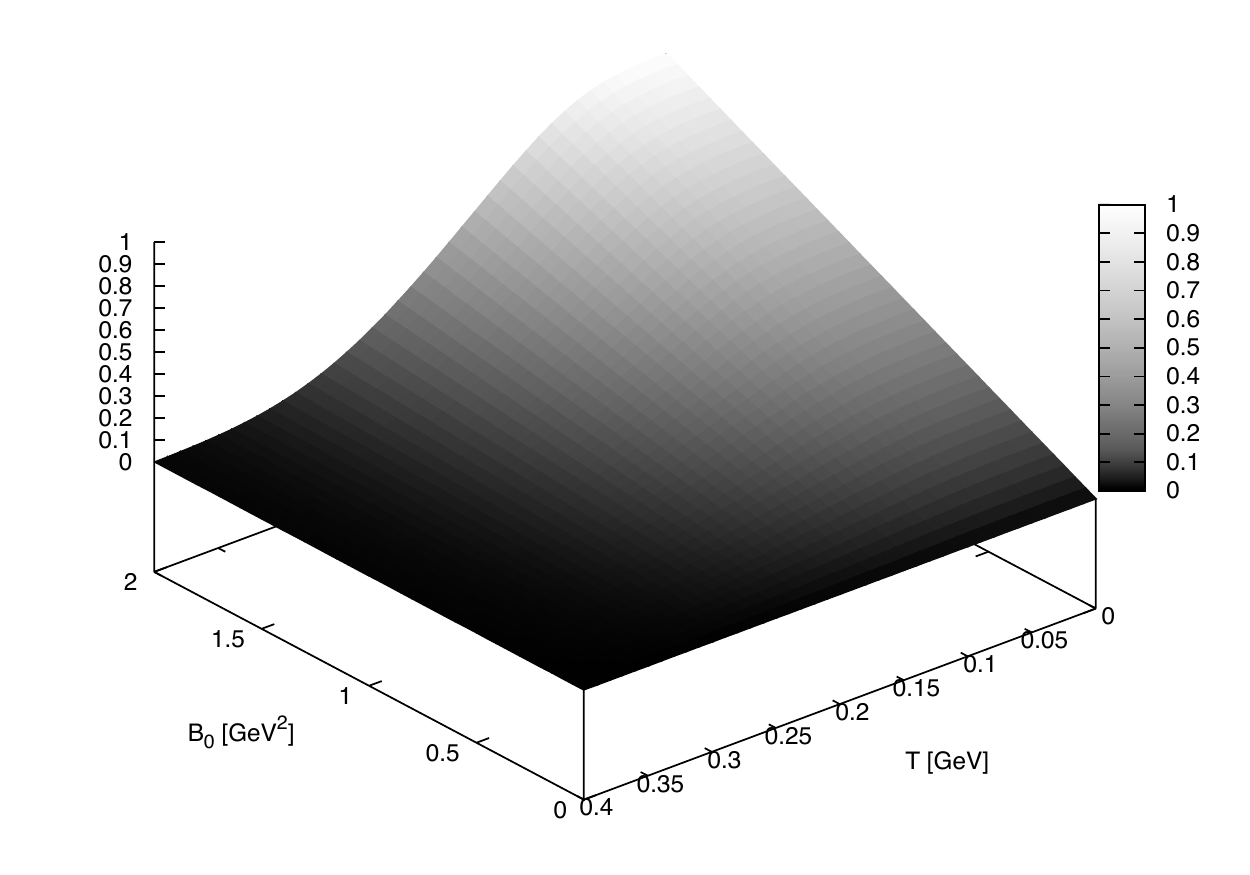}
\caption{Absolute value of the CME current with an appropriate normalization as a function of $T$ and $B_0$.}       
\label{FIG3}
\end{figure}

In this Section, we present our analytical and numerical results for the CME current, correlations, and charge separations via the present theoretical framework. First, we can extract a neat expression for the CME current which is parallel to $\bm{B}$ and equivalent to that given in Refs.~\cite{Fukushima:2009ft,Fukushima:2008xe}:
\begin{equation}
\label{eq:NA}
\mp\langle j_{3,4} \rangle_{F,\delta}
\approx\frac{1}{2\pi^{2}}(i\delta A_{4,3})B_{0}
=\frac{1}{2\pi^{2}}\mu_{\chi}B_{0},
\end{equation}
where we have assigned $i\delta A_{4,3}$ as $\mu_{\chi}$, which denotes the chiral chemical potential. In Eq.~(\ref{eq:NA}), we set $|\bm{B}|\equiv B_0$. Moreover, we can obtain the ratio between the longitudinal and transverse components of the CME current as:
\begin{equation}
\label{eq:JJJ}
\left|\frac{j_{\perp}}{j_{\parallel}}\right|=\frac{3}{2}\frac{|A_{2,1}|}{|A_{4,3}|}\delta.
\end{equation}
This expression tells us that the longitudinal component are much larger than the transverse one, because of that $CP$-violation strength $\delta\ll1$, assuming that $A_{1,2,3,4}$ are in a similar order. This observation is in good agreement with that from the LQCD simulation~\cite{Buividovich:2009wi}. In Fig.~\ref{FIG3}, we present the absolute value of the CME current with an appropriate normalization as a function of $T$ and $B_0$. As shown there, the CME current increases linearly with respect to $B_0$ as understood by Eq.~(\ref{eq:NA}), where it decreases as a function of $T$, indicating the diluting instanton ensemble, as discussed in the previous Section. Again, this typical observation is comparable with the LQCD simulation~\cite{Buividovich:2009wi}.

\begin{figure}[t]
\begin{tabular}{cc}
\includegraphics[width=8.5cm]{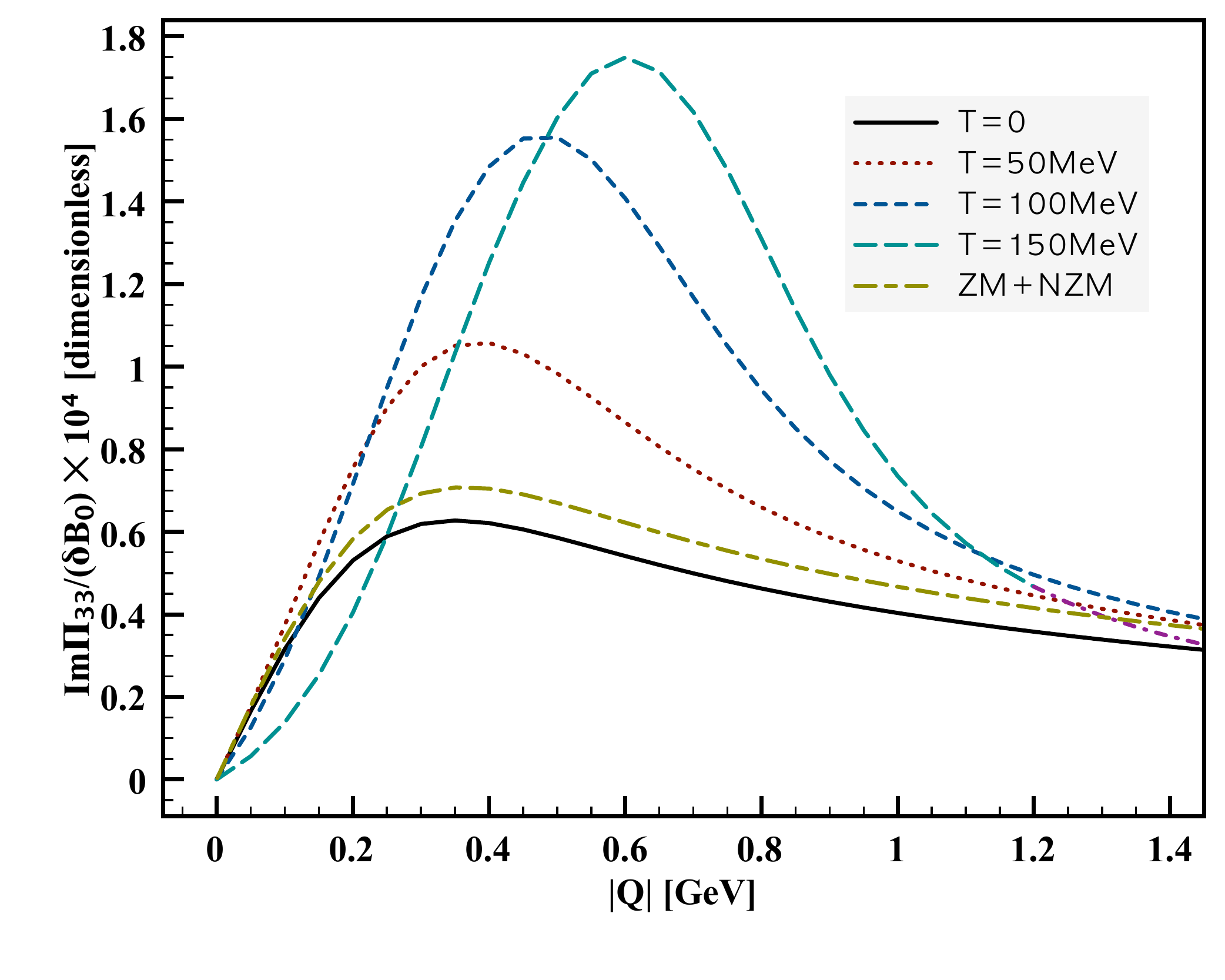}
\includegraphics[width=8.5cm]{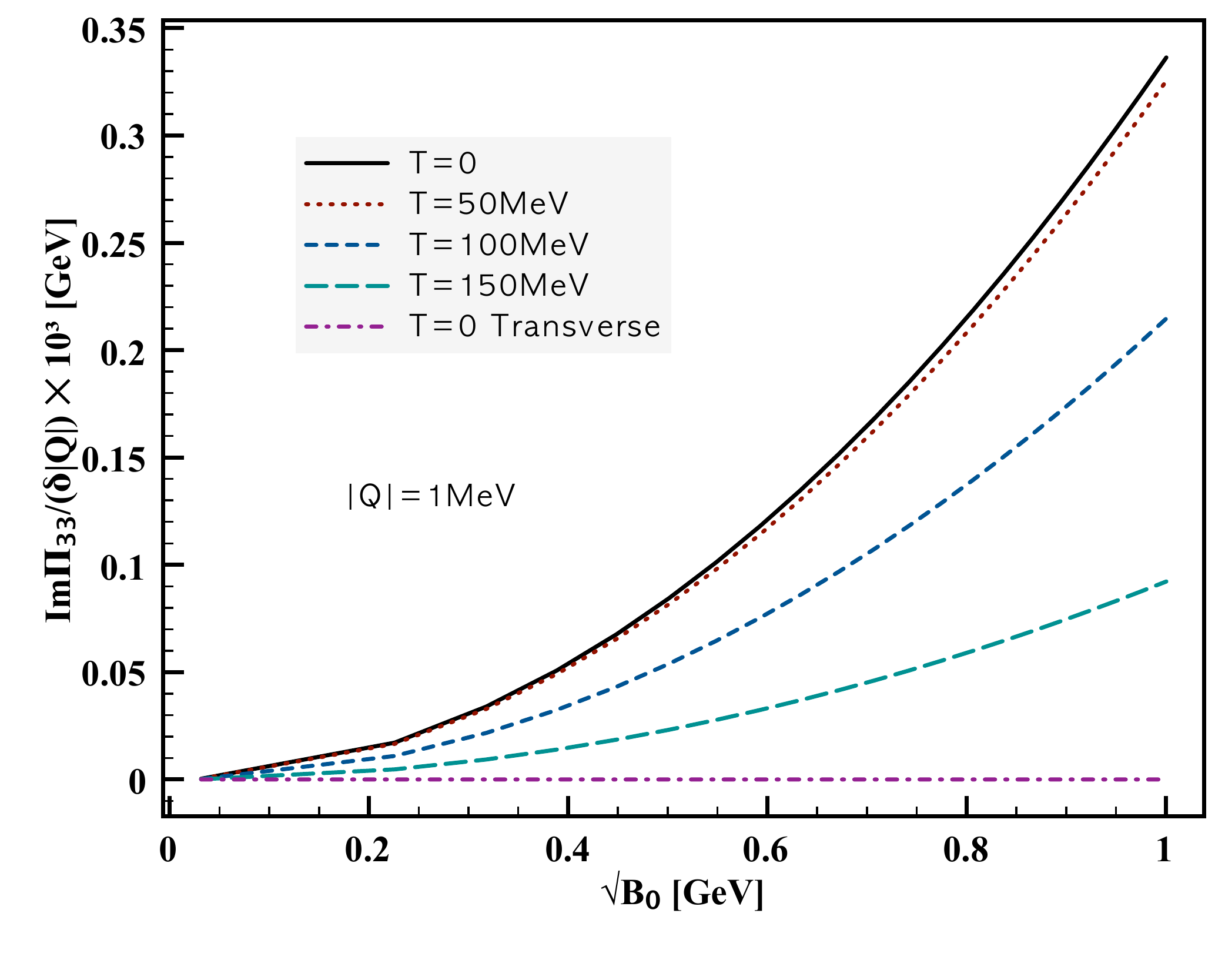}
\end{tabular}
\caption{Imaginary part of the CME current correlations as functions of the momentum transfer $Q$ (left) and the magnetic field $\sqrt{B_0}$ (right) with proper normalizations.}       
\label{FIG4n5}
\end{figure}

In Fig.~\ref{FIG4n5}, the imaginary part of the CME current correlations are given as functions of the momentum transfer $Q$ (left) and the magnetic field $\sqrt{B_0}$ (right) for different $T$ with proper normalizations. In the left panel of Fig.~\ref{FIG4n5}, as $T$ increases, the peak moves to higher $|Q|$ value, indicating a corresponding vector meson. It turns out that this shift of the peak is caused by the thermal mass of the quark in the quark propagator~\cite{Nam:2010nk}. This shift was also observed in the LQCD simulation~\cite{Buividovich:2010tn}. Being similarly to the CME current shown in Fig.~\ref{FIG3}, the correlations increase as $B_0$ does. Simultaneously, the strength of the correlations are decreasing with respect to $T$, due to the diluting instanton ensemble. Although the intercept values at $T=0$ are different, overall tendency shown here is compatible with that  in the LQCD simulation~\cite{Buividovich:2010tn}.

\begin{figure}[t]
\includegraphics[width=8.5cm]{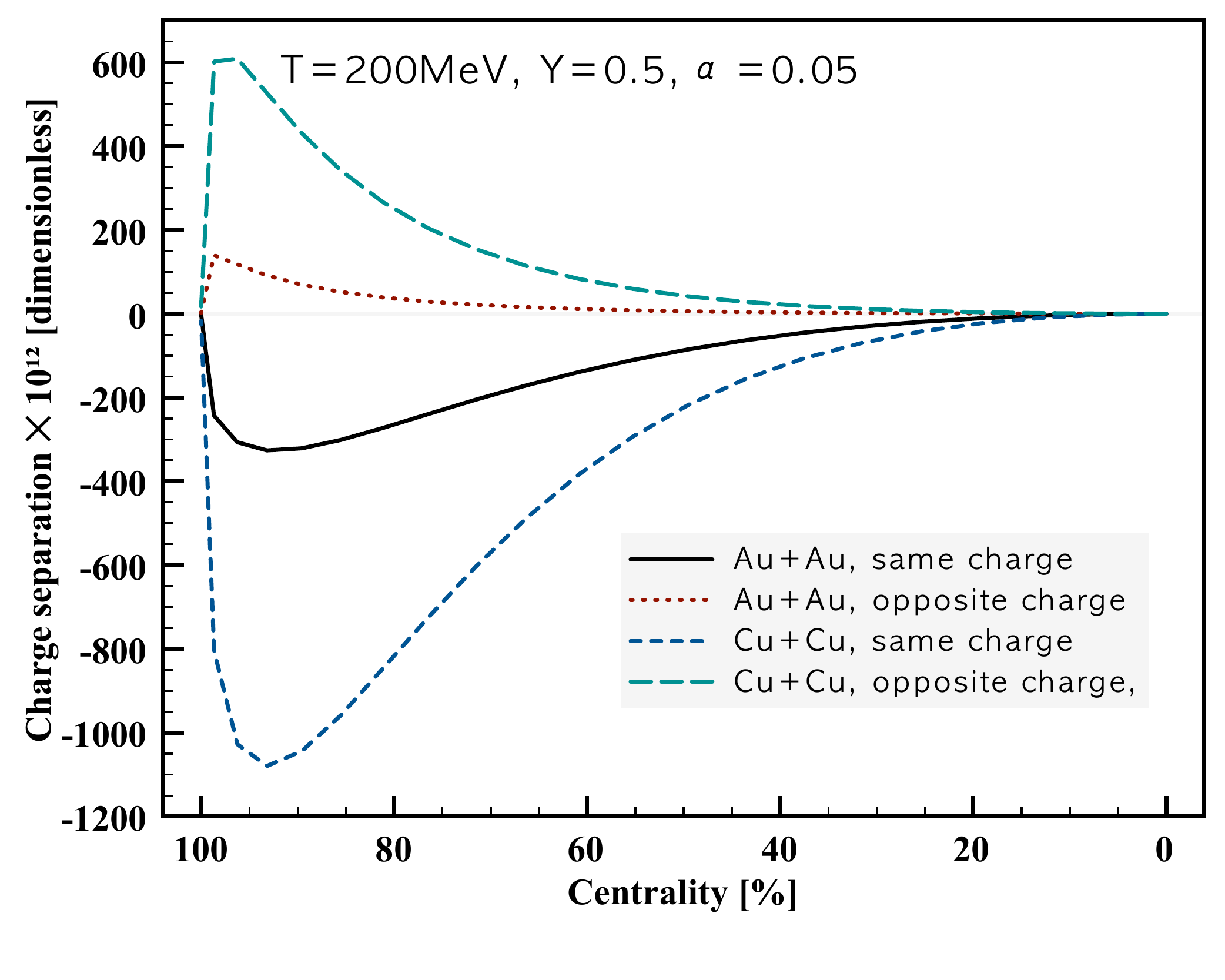}
\caption{Same and opposite charge separations for the Au$+$Au and Cu$+$Cu collisions as functions of centrality.}       
\label{FIG6}
\end{figure}

Finally, we compute the same and opposite charge separations, which are the experimental observable suggested in Refs.~\cite{Kharzeev:2004ey,Voloshin:2004vk}, denoted by
\begin{equation}
\label{eq:CHA}
\langle\langle \cos(\Delta\phi_{a}+\Delta\phi_{b})\rangle\rangle
\sim\langle J_{\parallel}\rangle^{2}_{\bm{B},\mu_{\chi}}
-\langle J_{\perp}\rangle^{2}_{\bm{B},\mu_{\chi}},
\end{equation}
where $\Delta\phi_{a,b}$ indicate the angles for the measured hadrons with the electric charges $a$ and $b$. Thus, $\langle\langle\cdots\rangle\rangle$ in Eq.~(\ref{eq:CHA}) stands for the correlation of the two hadron emissions, signaling the nontrivial event-by-event $P$ and $CP$ violations~\cite{Kharzeev:2004ey,Voloshin:2004vk,Kharzeev:2007jp,Voloshin:2008jx}. As understood in Eq.~(\ref{eq:CHA}), this physical quantity relates to the sum of the longitudinal and transverse components for the disconnected correlations. Although there are connected contributions, we ignored them due to thier smallness. Tunning some model parameters (see Ref.~\cite{Nam:2010nk} for more details), we can present the numerical results for those quantities at $T=200$ MeV in Fig.~\ref{FIG6}. We note that the present numerical results are in good agreement with the experimental data~\cite{:2009txa,:2009uh}. It is worth mentioning that the size effect, i.e. the charge separations are proportional to $1/N^2_\mathrm{nucl}$, plays an important role to reproduce the experimental data. This observation indicates that the probability for finding the $CP$-violating domain inside the QGP is proportional to $1/N^2_\mathrm{nucl}$.  

\section{Summary and outlook}
In this talk, we have reported the CME at the HIC experiments, employing the $T$-modified instanton-liquid model, considering the Harrington-Shepard. We computed the CME current and its correlations, using the linear Schwinger method to include the magnetic field in the present framework.  All the numerical results are compared with the model-independent approaches and LQCD simulations for the CME, and show qualitatively good agreement. An important observation in the present work is that the CME gets decreased as $T$ increases, due to the diluting instanton ensemble. Taking into account an appropriate magnetic potential configuration, screening effect, and size effect, we calculated the charge separations, which are the experimental observable measured by STAR collaboration at RHIC. It turns out that the size effect, corresponding to the probability to find the $CP$-violating domain inside the QGP,  plays an important role to reproduce the experimental data. More various physical quantities for the CME are under consideration, and related works will appear elsewhere. 

\section*{Acknowledgment}
This report was prepared for the talk, presented by S.i.N. at the $11$th Asia Pacific Physics Conference (APPC11), $14\sim18$ November 2010, Shanghai, China. S.i.N. is grateful to the organizers of the conference and the hospitality shown during his stay there. The authors appreciate the fruitful discussions with J.~W.~Chen and Y.~Kwon. The works of S.i.N. and B.G.Y. were  supported by the grant NRF-2010-0013279 from National Research Foundation (NRF) of Korea. The work of S.i.N. is also partially supported  by the grant NSC 98-2811-M-033-008 from National Science Council (NSC) of Taiwan. The work of C.W.K. was supported by the grant NSC 99-2112-M-033-004-MY3 from National Science Council (NSC) of Taiwan.

\end{document}